\documentclass[a4paper]{article}

\usepackage[ninepoint]{itgspeech2021}    
\usepackage{times}            
\usepackage[english]{babel}   
\usepackage[T1]{fontenc}      
\usepackage[sort&compress,numbers]{natbib}	
\usepackage{amsmath,amssymb}
\usepackage{graphicx}
\usepackage[colorlinks=false,pdfborder={0 0 0}]{hyperref}
\usepackage{units}
\usepackage{hyphenat}
\usepackage{balance} 
\title{An Objective Evaluation Framework for Pathological Speech Synthesis}

\author{Bence Mark Halpern$^{1,2,3}$, Julian Fritsch$^{4,5}$, Enno Hermann$^{4,5}$, Rob van Son$^{1,3}$ Odette Scharenborg$^{2}$, Mathew Magimai.-Doss$^{4}$}

\address{${}^{1}$University of Amsterdam, Amsterdam, The Netherlands \\
${}^{2}$Multimedia Computing Group, Delft University of Technology, Delft, The Netherlands \\
${}^{3}$Netherlands Cancer Institute, Amsterdam, The Netherlands \\ 
${}^{4}$Idiap Research Institute, Martigny, Switzerland \\
${}^{5}$\'Ecole polytechnique f\'ed\'erale de Lausanne (EPFL), Switzerland \\
Email: \texttt{\{b.halpern,r.v.son\}@nki.nl, \\ \{julian.fritsch, enno.hermann, mathew\}@idiap.ch, o.e.scharenborg@tudelft.nl}}

\begin{document}

\maketitle

\begin{abstract}
The development of pathological speech systems is currently hindered by the lack of a standardised objective evaluation framework. In this work, (1) we utilise existing detection and analysis techniques to propose a general framework for the consistent evaluation of synthetic pathological speech. This framework evaluates the voice quality and the intelligibility aspects of speech and is shown to be complementary using our experiments. (2) Using our proposed evaluation framework, we develop and test a dysarthric voice conversion system (VC) using CycleGAN-VC and a PSOLA-based speech rate modification technique. We show that the developed system is able to synthesise dysarthric speech with different levels of speech intelligibility.

\end{abstract}

\section{Introduction}
\label{sec:intro}

In recent years, there has been a growing interest in pathological speech processing \cite{gupta2016pathological}. It is a highly challenging area, as our understanding of speech is largely limited to ``typical" and unimpaired speech. The advances made in detection and analysis of pathological speech are continuously improving our understanding of it. These efforts could be further accelerated with pathological speech synthesis - as speech processing has advanced through the synergy between "analysis" and "synthesis". For instance, if one considers the message component in the speech signal, automatic speech recognition (ASR) can be regarded as analysis and text-to-speech synthesis (TTS) can be regarded as synthesis. Such understandings have 
benefited both ASR and TTS and have driven innovation, e.g., \cite{tokuda_vlbrc_hmm_icassp98, leeVLBRC_2001, Dines_JSTSP_2010, tjandra2017listening}. 

Another motivation for the development of pathological speech synthesis is that it could assist in informed decision making for the medical conditions at the root of the pathological speech. For instance, oral cancer surgery results in changes to a speaker's voice. Availability of a synthesis model that can generate how the voice could sound after surgery could help the patients and clinicians to make informed decisions about the surgery and alleviate stress of the patients \cite{Epstein1999}.

Pathological speech synthesis is a non-trivial task due to two main challenges. Firstly, state-of-the-art methods for TTS/voice conversion (VC) often rely upon the availability of a large amount of speech data from many different speakers and linguistic resources. In that sense, pathological TTS/VC is inherently under-resourced, as the collection of pathological speech is time-consuming and is governed by medical research ethical aspects.

Secondly, a crucial aspects in the development of TTS/VC systems is the evaluation of the naturalness of the synthesised speech. This is typically done by asking a group of naive human listeners to rate the synthesised speech on its naturalness \cite{wu2019blizzard}. The most common example of such an evaluation by naive listeners is the mean opinion score (MOS). MOS is a good evaluation measure for TTS of typical speech where the only deterioration in the speech is due to the decrease in naturalness. However, it is not appropriate for pathological speech, for several reasons. Firstly, a pathological VC system that receives better MOS scores than the reference pathological speech is likely not able to capture the characteristics of the pathological speech. Secondly, on the reverse, a VC system mimicking the pathology, albeit in an exaggerated manner, would likely produce a MOS score that is a lot lower than that of the reference. Thirdly, it is unknown whether non-expert listeners have sufficient expertise to evaluate pathological speech. Fourthly, involving speech-language pathologists for evaluation at every stage of development would make the evaluation process not only costly but it is also not certain that they can distinguish the unnaturalness of the synthesised speech from the characteristics of the pathology.

To alleviate this problem, the present paper
\begin{itemize}
    \item[(a)] proposes a new framework, based on the synthesis-analysis approach for ASR, for the consistent evaluation of limited-resource, synthesised pathological speech based on a combination of pathological speech detection and analysis methods.
    \item[(b)] and, to that end, investigates a CycleGAN-based approach to convert healthy control speech to pathological speech of different levels of intelligibility.
    
\end{itemize}

Specifically, in this paper, we focus on the synthesis of dysarthric speech and its evaluation through pathological speech detection (SD) and automatic intelligibility assessment. We demonstrate the viability of the proposed framework through an investigation on the UASpeech corpus. 

The paper is organised as follows: the UASpeech database used for training and evaluation is described in Section \ref{sec:database}. The employed CycleGAN-VC method is described in Section \ref{sec:methods}.   We present the proposed objective evaluation framework  in Section \ref{sec:experiments} and the evaluation results in Section \ref{sec:results}. Speech samples and a demo of the system is available online.\footnote{\url{https://karkirowle.github.io/publication/itg-2021-dysarthric-vc}}

\section{Database}
\label{sec:database}

For all experiments, we use the UASpeech database \cite{kim2008dysarthric}, which contains 14 dysarthric speakers, 10 male and 4 female, and 455 unique isolated word recordings for each speaker. The speakers have received an intelligibility rating in the database (high, mid, low, very low) based on their average scores in a word transcription task rated by five non-expert listeners (subjective intelligibility score). In our experiments, low and very low intelligibility speakers are grouped together to increase statistical power. In addition, the database contains 9 male and 4 female unimpaired, control speakers. All recordings were done with multiple microphones and repetitions, however, only microphone 5 and the first repetition of each word was used in our experiments. We refer to the healthy speakers' utterances from the database as ground truth (GT) healthy, and the dysarthric utterances as ground truth dysarthric from now on.

\begin{figure*}[!htb]
    \centering
    \includegraphics[width=\textwidth]{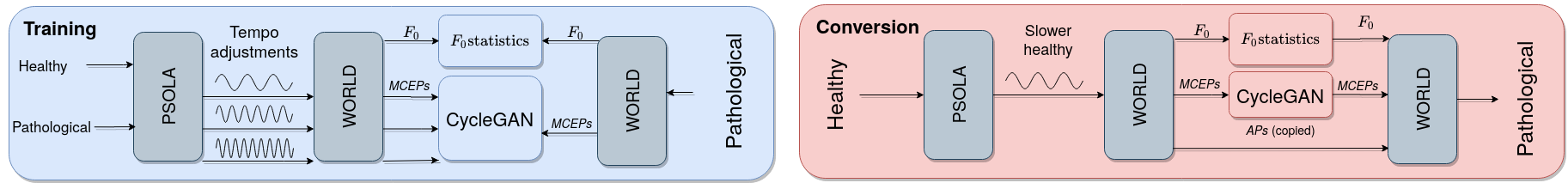}
    \caption{Training (blue) and conversion (red) setup for dysarthric voice conversion.}
    \label{fig:training}
\end{figure*}

\section{Proposed conversion system}
\label{sec:methods}

We train a voice conversion model for each dysarthric speaker in the database, using CM08 as a male source speaker and CF04 as a female source speaker. 
This means that -- contrary to traditional VC -- we are not only converting speaker traits, we are also aiming to change the intelligibility of the speech. The overall experimental setup is illustrated in Figure \ref{fig:training}. In the following sections, we will explain the techniques used in detail.

\subsection{Speech rate modification and analysis}

VC systems (including CycleGAN-VC) do not change the speech rate, but for pathological speech applications, it is essential that VC systems convert this aspect of the speech: e.g., in the case of dysarthric speech, speech rate has been shown to be a reliable predictor of speech severity \cite{vachhani2018data}. Therefore, prior to conversion, the tempo of the source speech is adjusted to the dysarthric speaker's tempo using the pitch synchronous overlap-add algorithm (PSOLA) \cite{moulines1990pitch} from Praat \cite{boersma2011praat}. We've run preliminary tempo modification experiments with WSOLA (\texttt{sox} implementation) and phase vocoding (librosa \cite{brian_mcfee_2020_3955228} implementation) techniques too, but we found that PSOLA results in the most natural sounding speech. We would like to emphasise that, even though the target speaker's tempo is not generally known beforehand, for all practical purposes, the expected speech rate can be estimated if the severity is known. Such estimation has been already done in various works, such as \cite{xiong2019phonetic, vachhani2018data, 10.1007/978-3-319-43958-7_44}.

Furthermore, a tempo-based data augmentation scheme is used to increase the dataset size in the CycleGAN-VC training, which means that the same speech signal is fed into the neural networks three times with three different speech rates, modified using the pitch synchronous overlap-add algorithm from Praat. The first version is the original unadjusted version of the speech signal, spoken at the rate of the control speaker. The second version is adjusted to the speech rate of the dysarthric speaker. The last version is adjusted to a rate exactly halfway in between the original unadjusted and the dysarthric adjusted. This tempo-based data augmentation scheme increases the size of the training set three-fold.  We refer to the dataset obtained after the augmentation as the augmented database, and the speech adjusted to dysarthric speech rate as adjusted speech. Subsequently, the augmented database is preprocessed using the WORLD vocoder \cite{morise2016world} to obtain the Mel-generalised cepstrum (MCEP), the pitch ($F_0$) and the aperiodicities (AP). 

\subsection{CycleGAN-VC training and conversion}
\label{sec:acoustic_modeling}

Previous studies have found that CycleGAN-VC-based systems can improve the speech intelligibility of a dysarthric signal \cite{yang2020improving,purohit2020intelligibility}. For this reason, we are interested if a CycleGAN-VC based system can also model the deterioration of speech intelligibility too. An additional benefit of using the CycleGAN-VC is that the system is known to work with small amount (approximately 10 mins) of data, which can be non-parallel. This makes CycleGAN-VC lucrative for pathological speech applications which are inherently low-resourced.

A CycleGAN-VC system is trained with the augmented database \cite{Kaneko2018CycleGANVCNV}\footnote{The implementation used was \url{https://github.com/leimao/Voice_Converter_CycleGAN}} with the same hyperparameters and values  (as most hyperparameters are directly adapted from the original CycleGAN paper) 
for 1000 epochs \cite{zhu2017unpaired}. During conversion of the healthy speech to dysarthric speech, only the healthy tempo-adjusted speech is used. The MCEPs are then converted using the trained model. The APs are copied. During training,  the mean and standard deviation of $F_{0}$ values are calculated for both speakers (source and target). The source $F_0$ is mean-std normalised to the source speaker's $F_0$ distribution and mean-std unnormalised to the target speaker's $F_{0}$ distribution.

\section{Objective evaluation framework}
\label{sec:experiments}

Pathological speech deviates from healthy speech along several different dimensions. Therefore, the proposed system has four parts, each of which evaluates the pathological speech signal generated by the VC from a different angle, together yielding a consistent evaluation measure: a pathological speech detector (Section \ref{sec:ltas-lasso-sd}), a voice quality measure (Section \ref{sec:ltas-skl}), an intelligibility assessment tool (Section \ref{sec:ppg-dtw}) and an ASR system (Section \ref{sec:asr}).

Before the evaluation experiments, voice activity detection (VAD) is performed using Praat to avoid analysis of non-speech cues \cite{boersma2011praat}. This VAD is manually checked for quality. Ideally, VAD would be done before training, however, our experiments found that in this case the CycleGAN-VC does not converge well due to the short duration of the utterances.

\subsection{Voice quality measure: LTAS-LASSO-SD}
\label{sec:ltas-lasso-sd}

Long time average spectrum (LTAS) has been successfully used as a voice quality measure \cite{master2006long} and predictor for various speech pathologies \cite{halpern2020detecting,smith2014long}. Thus, an LTAS based pathological speech detector (SD) will be trained on detecting dysarthric speech using the voice quality cues provided by the LTAS-feature, assessed for generalisation using the GT samples, and then used to evaluate our converted (VC) samples. 

\begin{figure*}[t]
    \centering
    \includegraphics[width=\textwidth]{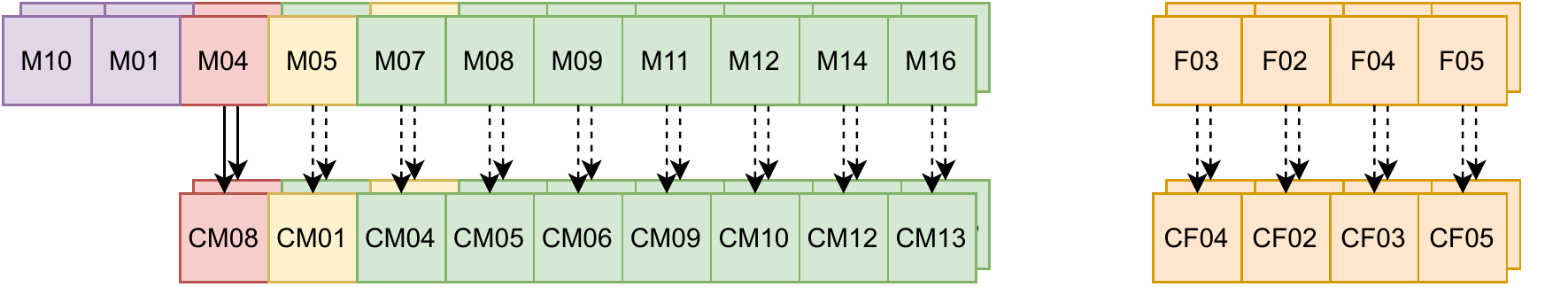}
    \caption{Leave-one-out validation scheme for LTAS-LASSO-SD. Purple colour shows the excluded speakers. The conversion pair (red) is excluded from the evaluation, and in each fold a different speaker pair (yellow) is left out. The green colour shows the speakers that we are training on. This is repeated for the two genders separately.}
    \label{fig:experiment_design}
\end{figure*}

The leave-one-out validation scheme used for the LTAS-LASSO-SD is illustrated in Figure \ref{fig:experiment_design}. The main factors considered during the design of this validation scheme were that (1) the speaker who is evaluated should not be included in the training data (2) each speaker should be mapped with a single control speaker so that everyone is only left out during its own evaluation round and that the number of control and pathological speakers remain changed.
Two male speakers had no control pairs, so we excluded one with very high intelligbility (M10) and one with very low (M01). The weakness of this experiment design is that male speakers are overrepresented in the dataset. However, the alternative would be one single speaker per gender per severity, which we found an inferior design. 

First, an LTAS is extracted based on a 512-point FFT with Hann window and 128 sample frame shift with librosa \cite{brian_mcfee_2020_3955228} for the control, VC and original dysarthric speech.

The chosen detector model was a LASSO due to its automatic feature selection property. The sparsity penalty was \( \alpha = 10^{-5} \). The model was trained for a maximum of 1000 iterations. The VC samples are then evaluated on these detector models in terms of accuracy. Furthermore, we investigate to what extent voice quality is influenced by the subjective intelligibility of utterances (taken from the UASpeech database) by performing a Pearson's correlation ($r$) between the subjective intelligibility scores and mean detection scores of the VC utterances.

\subsection{Intelligibility decrease measure: LTAS-SKL measure}
\label{sec:ltas-skl}

The idea of the LTAS-SKL measure is the following: each word of a reference speaker is compared to the same word of all other speakers (control and pathological; VC and GT). The comparison is done on the distribution of LTAS energy bins using the symmetric Kullback-Leibler (SKL) divergence. The speakers are then grouped based on intelligibility and the results visualised using a box plot. This way we ensure that the difference in the utterances can only be due to the individual differences in speakers (spread) and due to the differences in intelligibility of the healthy ground-truth speech and the generated pathological speech (median). Here, a normalised LTAS with a 1024-point FFT with Hann window and 256 sample frame shift is extracted. 
A t-test is used to check the significance  of the difference in intelligibility.

\subsection{PPG-DTW-based utterance verification}
\label{sec:ppg-dtw}

The phonetic posteriorgram (PPG)-dynamic time warping (DTW) method~\cite{Fritsch_Idiap-RR-01-2021} evaluates pathological utterances by matching them to a healthy reference utterance of the same word and aligning their PPG sequences with DTW. To verify an utterance, the DTW matching score is converted into a probability $P_c$, by plugging it into a logistic function, $c$ denoting the control class; an utterance is hence verified correctly if $P_c \ge 0.5$. A speaker's intelligibility is then estimated in terms of the percentage correctly verified utterances, just as in human listening tests.

For our experiments, we used an SKL local cost function in the DTW implementation and posterior sequences of 45 context-independent phones (PPG). Spearman's $\rho$ and Pearson's correlation $r$ analyses between the intelligibility estimates of the dysarthric speech GT and the VC speech, show that the method performs at $\rho=.965$, $r=.854$, which is comparable to other state-of-the-art methods such as \cite{janbakhshi2019pathological}. The evaluation is repeated on the WORLD vocoder resynthesised GT dysarthric signals to see if vocoding affects the intelligibility estimates.

\subsection{ASR based intelligibility evaluation}
\label{sec:asr}

As an additional objective intelligibility measure, we use the word error rate
(WER) of an ASR system trained only on unimpaired control speakers. We examine
whether the VC speech gives comparable results to the GT dysarthric data to show that the VC system learns to model the lower intelligibility of dysarthric speech.

We trained acoustic models with the Kaldi ASR toolkit~\cite{Povey2011} on block 2
and 3 of the UASpeech corpus, which contain distinct sets of words from block 1 that was used for VC. We
trained the acoustic models on the data from all microphones and all control speakers
except the 2~source speakers for VC (CM08, CF04), based on an existing Kaldi recipe~\cite{Anonymous2020} and decode with a
unigram grammar that includes all words in the UASpeech corpus. To verify that the results are not specific to a single type of model, we compare both
subspace Gaussian mixture models (SGMMs)~\cite{Povey2010} and neural networks
trained with the sequence-discriminative lattice-free maximum mutual information
(LF-MMI) loss~\cite{Povey2016}, the state-of-the-art methods in the GMM and hybrid hidden Markov model/deep neural network ASR paradigms, respectively.

\begin{figure*}[ht]
    \centering
    \includegraphics[width=\textwidth]{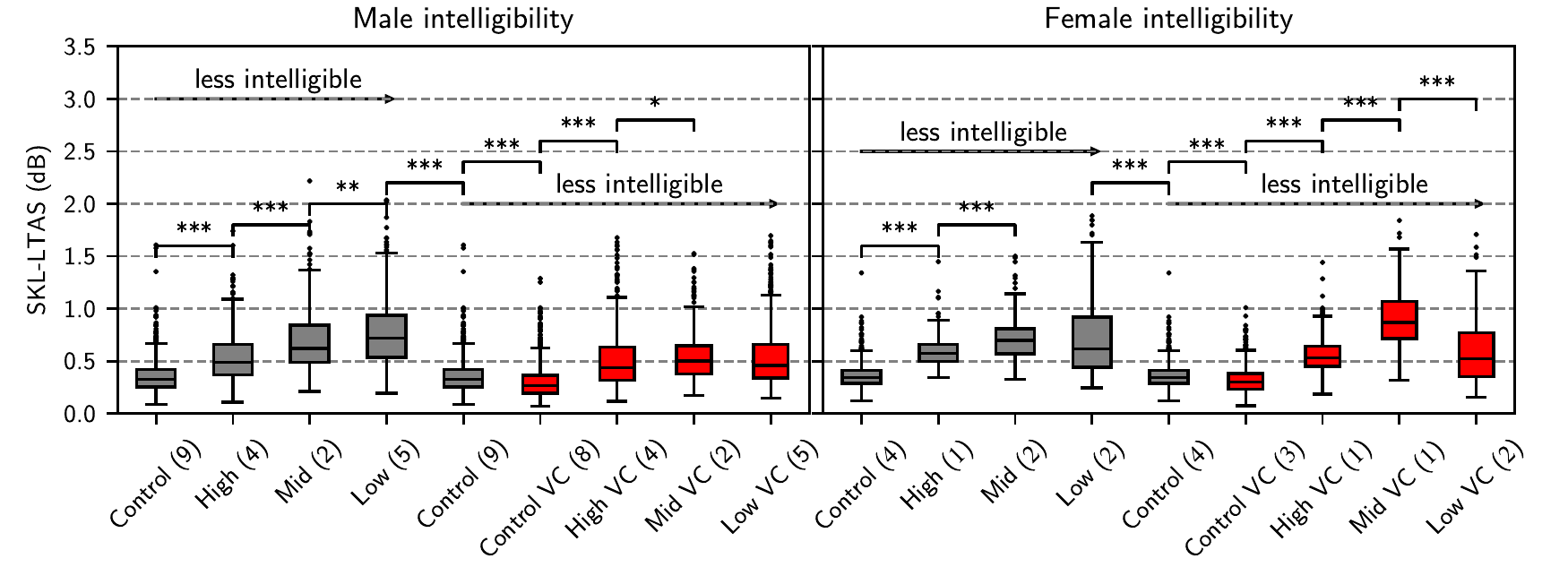}
    \caption{Result of the LTAS-SKL analysis. A grey interquartile range (IQR) indicates original samples, a red IQR indicates VC samples. The center line indicates the median. (***) $p < .001$ (**) $p < .01$ (*) $p < .05$, otherwise not significant. Number of speakers are written in parentheses. The ground truth control speakers of are plotted twice on each panel for easier comparison.}
    \label{fig:ltas_experiment_2}
\end{figure*}

\section{Results and discussion}
\label{sec:results}

\subsection{Voice quality measure: LTAS-LASSO-SD}

The LTAS-LASSO-SD detectors perform well on the held-out GT speakers, with the best held-out speaker having an accuracy of $93.15\% \pm 9.75 $ (M14), the worst held-out speaker performing with a $90.84\% \pm 10.1$ (M05). This means that the proposed evaluation framework is able to detect the voice quality of dysarthric speech. On the VC samples the best held-out speaker accuracy is $100\% \pm 0$ (multiple speakers, i.e., M16, F02, F03, F04, F05) and the worst accuracy is $92.07\% \pm  6.05$ (M11). This means that the proposed VC system successfully mimics the voice quality aspects of dysarthric speech.

There is a linear relationship between the male speakers' detection scores and the subjective intelligibility scores from UASpeech (male; $r = -0.85$, $ \text{p-value} = 0.003$), which shows that the intelligibility aspects of the synthesised dysarthric speech are closely related to the voice quality aspects in case of the male speakers. This relationship does not exist in the case of females (female; $r = -0.02$, $ \text{p-value} = 0.98$). This means, while the voice quality aspects can be heavily influenced by the intelligibility of the speakers in the case of male speakers, the proposed voice quality measure should be used in complement to the other, intelligibility-based measures.

\subsection{Intelligibility decrease measure: LTAS-SKL}

In Figure \ref{fig:ltas_experiment_2}, it can be seen that both for the GT (gray) and the VC (red) speakers the median of the LTAS increases (SKL-LTAS dB) as the intelligibility decreases. This increase in the median is significant, except for mid-low VC male and mid-low GT female. This means that VC performs well overall, however it is not that well able to generate dysarthric speech that differentiates between mid and low intelligibility. 

\subsection{PPG-DTW-based utterance verification}

Table \ref{tab:results} shows that the correlations between the subjective and objective intelligibility scores are high, and are significant for all male speakers, meaning that the proposed VC model successfully captures characteristics of dysarthric speech that influence intelligibility for male speakers. The results also show high correlation for female speakers, though not significant, which might be due to the lower number of female speakers. Additionally, it can be seen that overall there is a correlation gap between the WORLD all (copy synthesis) and VC all (conversion). This indicates that conversion performance is not limited by the vocoder, but rather by the used model.

\begin{table}[!htb]

  \label{tab:results}
  \centering
  \begin{tabular}{l|cccc}
	\hline
 Data & $\rho$ &  p-value & $r$ & p-value  \\
 \hline
 \hline
  WORLD all & .954 & *** & .836 & *** \\
  VC all & .448 & *  & .481 & * \\
   \hline
    WORLD male & .940 & *** & .807 & ** \\ 
 VC male & .735 & **  &  .594 & * \\
 \hline
  WORLD female & 1.0 & *** & .957 & *  \\  
 VC female & .774 & 0.229  & .720 & 0.279 \\   
 \hline
  \end{tabular}
 
    \caption{Pearson's and Spearman's correlation between subjective intelligibility score (from UASpeech) and predicted intelligibility by PPG-DTW. (*) $p < .1$ (**) $p<.05$.}
\end{table}

\begin{figure}[ht!]
\centering
  \includegraphics[width=\columnwidth]{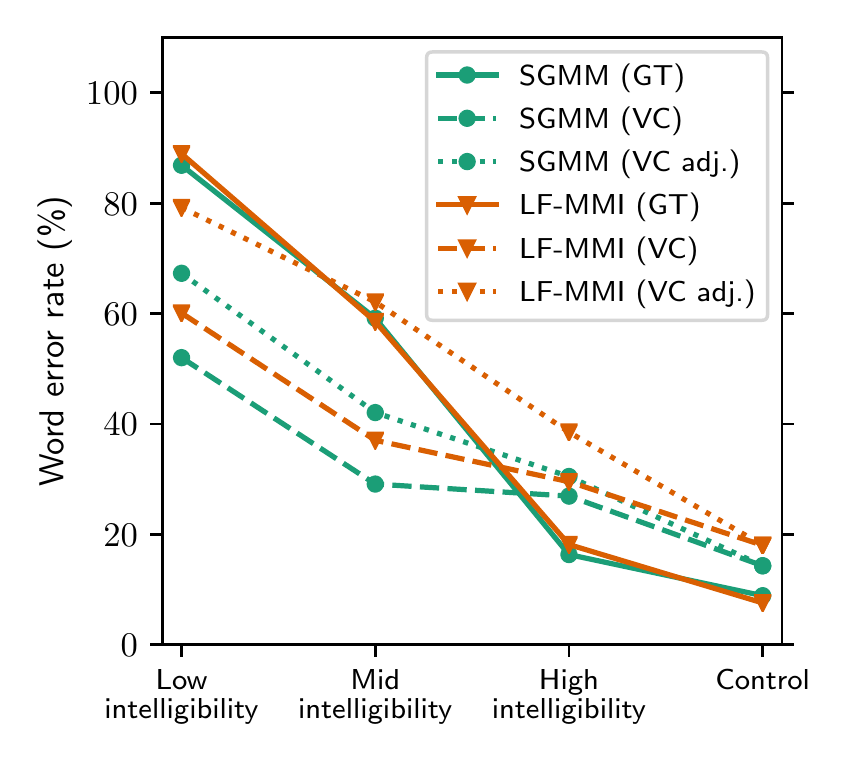}
  \caption{WER results grouped by intelligibility of the target speakers, comparing GT
    and VC speech using original and adjusted tempo (VC adj).}
  \label{fig:asr-results}
\end{figure}

\subsection{ASR-based intelligibility evaluation}
\label{sec:asr-results}

Figure~\ref{fig:asr-results} shows the WER obtained with the two ASR
systems for the low to highly intelligible dysarthric and the control speakers. As expected, the
GT speech data (solid lines) exhibits a steep drop in error rates as intelligibility
increases. The VC speech mirrors this decrease. When the speech rate is adjusted (dotted lines), the WERs resemble the ground
truth more closely. This is consistent with previous work that has shown slower speaking rates of dysarthric speech to harm ASR
performance~\cite{rudzicz2013adjusting}. However, the same pattern of fewer errors when intelligibility is higher is still apparent without tempo adjustment (dashed lines). This shows that the VC system
learns to manipulate the speech in such a way that makes it harder for an ASR
system trained only on control speech to recognise it correctly. Both when synthesising healty and dysarthric speech,
we expect the VC speech to perform slightly worse than the GT
because of the added noise introduced by the VC, which does not match the acoustic conditions of the ASR
training data. We observe the same broad pattern for SGMM (green) and LF-MMI (orange) models, indicating that our analysis generalises across different types of acoustic models.

\section{Conclusion}

In this paper, we proposed an objective evaluation framework for pathological speech synthesis. The proposed framework (1) captures the voice quality aspects of pathological speech, and consistently shows the decrease in intelligibility of synthesised dysarthric speech compared to healthy speech using the (2) LTAS-SKL (3) PPG-DTW (4) and an ASR-based methods. Simultaneously, we have developed and evaluated a proof-of-concept CycleGAN-VC system which (1) mimics the voice quality of dysarthric speech, (2,3,4) can exhibit different levels of speech intelligibility (4) and shows that this decrease in intelligibility is independent of the decrease in speech rate.

\section{Acknowledgements}
This  project  has  received  funding  from  the  EU H2020 research and innovation programme under MSCA grant  agreement  No  766287 (TAPAS).   The  Department of Head and Neck Oncology and surgery of the Netherlands Cancer Institute receives a research grant from Atos Medical (H\"orby, Sweden), which contributes to the existing infratructure for quality of life research.

\newpage
\small
\bibliographystyle{ieeetr}
\bibliography{refs}


\end{document}